\documentclass[conference]{IEEEtran}
\IEEEoverridecommandlockouts
\usepackage{cite}
\usepackage{amsmath,amssymb,amsfonts}
\usepackage{algorithmic}
\usepackage{graphicx}
\usepackage{textcomp}
\usepackage{xcolor}
\def\BibTeX{{\rm B\kern-.05em{\sc i\kern-.025em b}\kern-.08em
    T\kern-.1667em\lower.7ex\hbox{E}\kern-.125emX}}
\begin{document}

\title{An Investigation into the Performances of the State-of-the-art Machine Learning Approaches for Various Cyber-attack Detection: A Survey}

\author{\IEEEauthorblockN{1\textsuperscript{st} Tosin Ige}
\IEEEauthorblockA{\textit{dept. of Computer Science} \\
\textit{The University of Texas at El Paso}\\
Texas, USA \\
toige@miners.utep.edu}
\and
\IEEEauthorblockN{2\textsuperscript{nd} Christophet Kiekintveld}
\IEEEauthorblockA{\textit{dept. of Computer Science} \\
\textit{The University of Texas at El Paso}\\
Texas, USA \\
cdkiekintveld@utep.edu}
\and
\IEEEauthorblockN{3\textsuperscript{rd} Aritran Piplai}
\IEEEauthorblockA{\textit{dept. of Computer Science} \\
\textit{The University of Texas at El Paso}\\
Texas, USA \\
apiplai@utep.edu}

}

\maketitle

\begin{abstract}
In this research, we analyzed the suitability of each of the current
state-of-the-art machine learning models for various cyberattack detection from the past
5 years with a major emphasis on the most recent works for comparative study to identify the
knowledge gap where work is still needed to be done with regard to detection of each category of cyberattack.
We also reviewed the suitability, effeciency and limitations of recent research on state-of-the-art classifiers and novel frameworks in the detection of differnet cyberattacks. Our result shows the need for; further research and exploration on machine learning approach for the detection of drive-by download attacks, an investigation into the mix performance of Naive Bayes to identify possible research direction on improvement to existing state-of-the-art Naive Bayes classifier, we also identify that current machine learning approach to the detection of SQLi attack cannot detect an already compromised database with SQLi attack signifying another possible future research direction.
\end{abstract}


\section{Introduction}
The fact that modern systems are not perfect guarantees that there will always be vulnerabilities no matter how small which could be exploited by an attacker to have unauthorized access which will enable him to violate security policy. Every modern system has vulnerabilities that could be exploited because exploitation could be developed for any system whose vulnerabilities could be described, hence, attacks are easily developed the moment such vulnerabilities are found. It is for this reason that finding vulnerabilities that are previously undiscovered as one of the proven ways to be a hacker elite is a strong cybersecurity culture. At the same time, an exploit is an attack through the vulnerability of a computer system with the purpose of causing either denial-of-service (DoS), install malware such as ransomware, Trojan horses, worms, spyware and so on. The result of a successful attack is what leads to security breach which is an unauthorized access to entity in the cyberspace, this often result in loss of confidentiality, integrity, or availability of data and information, as the attacker is able to remove or manipulate sensitive information.

On the other hand, modern attacker have developed a sophisticated social engineering technique by the use of low level of technology attack such as impersonation, bribes, lies, tricks, threats, and blackmail in order to compromise computer system. Social engineering usually relies on trickery for information gathering and the aim is to manipulate people to perform action(s) which will lead to the attacker getting confidential information of the person or organization. Phishing attack falls into this category of attack because the attacker use trick that can eventually lead the victim into divulging sensitive and personal information which attacker can use to gain access to server, compromise organization system, or commit various cyber crime which includes but not limited to business e-mail compromise (BEC), phishing, malware attack, denial of service (DDoS) attack, Eavesdropping Attacks, Ransomware attack and so on.

In order to secure computer and information systems from attacker taken advantage of vulnerabilities in system to commit cybercrime, several methods had been adopted for earlier detection of vulnerabilities as well as quick or real time detection of comprise in computer information system space to improve security around computer and information system. In all the methods machine learning had been the most effective methods in securing system with capability ranging from early detection of software vulnerabilities to real-time detection of ongoing compromise in a system. Each of the existing machine learning classifier models depends on different algorithms such as Support Vector Machine (SVM), Logistic Regression (LR), Naïve Bayes classifier, deep learning based, decision tree, random forest, XGBoost and so on,   for which they are suitable for different kind of cybersecurity-related classification tasks. 
Having observed the under-performance of Naïve Bayes variants in comparison with another suitable classifier for the same cybersecurity-related tasks such as phishing detection, anomaly detection in network intrusion, software vulnerability detection, malware detection, and so on, several work had been done by to improve the performance of Naïve Bayes classifier by other researchers. In this review paper, we analyzed both the performance and result of various proposals from the past 10 years to address the underperformance of Naïve Bayes-based classifier to better understand the current state-of-art of  Bayesian-based classifier.

\begin{table*}
\centering
\caption{Limitations of current state-of-the-art Approaches for cyberattack detection}
\begin{tabular}{|p{0.7in}|p{0.8in}|p{1.5in}|p{1.5in}|p{1.8in}|}
 \hline
 Author & Dataset  & Research Summary & Method/Algorithm & Limitation  \\
 \hline

 Mayank Agarwal et al.\cite{agarwal2016machine}  & Internally generated dataset  &Proposal of IDS to detect the existence of flooding-based DoS attacks in a Wi-Fi network  & Naive Bayes, Bayes Net, Ridor, ADTree & 1. The dataset used is not publicly available \newline 2. An unknown number of records in the generated dataset   \\
\hline
 Ali Bin et al. (2022) \cite{sultan2022man}  & Kali Linux distribution having 3855 attacks & evaluation of machine learning classifiers for the detection of man-in-the-middle attacks  & Decision Tree \newline Naive Bayes \newline K-Nearest Neighbor(KNN) \newline Artificial Neural Network \newline XGBoost & 1. The effect of feature variation on the classifier was not investigated  \newline 2. The result obtained is limited to only one dataset   \\
\hline

Muhanna Saed and Ahamed Aljuhani (2022) \cite{saed2022detection}   & Wi-Fi network benchmark dataset  &Proposal machine learning techniques both to detect and identify Man-In-The-Middle attacks on a wireless communication network  & LSTM, Support Vector Machine, Random Forest & 1. proposed model likely to be bias due to imbalance dataset \newline 2. Imbalance dataset as attack 2 packet is above 12000, attack 1 is less than 500, while normal packet is about 2000.      \\
\hline

 Kamal Omari (2023) \cite{Omari2023}  &  UCI phishing domains dataset  & an investigation into the performances and efficiency of Logistic Regression, KNN, SVM, Naive Bayes, Decision Tree, Random Forest, and Gradient Boosting for phishing detection task  & naive bayes \newline KNN \newline SVM \newline Random Forest \newline Gradient Boost \newline Decision Tree \newline Logistic Regression & 1. While Random Forest have a good accuracy of 97.1\%, its complexity of re-generating tree still remains \newline 2. Heavy reliance on URL attributes which means the model remains vulnerable     \\ 
\hline

Prince Roy et al. (2022) \cite{sultan2022man}  & Kaggle SQL Injection attack dataset & evaluation of machine learning classifiers for the detection of SQL Injection attacks  & Logistic Regression  \newline Naive Bayes \newline XGBoost \newline Random Forest \newline Adaboost & 1. Limited only to SQL injection, does not address other web application vulnerabilities  \\
\hline

Ann Zeky et al (2023) \cite{magdacy2023detect}  & Internally generated dataset  & proposal of extraction based naive Bayes robust model for phishing detection with emphasis on the combination of webpage content and URL feature analysis & naive Bayes \newline URL analyzation \newline webpage content extraction  & 1. problem of bayesian poisoning was not addressed and so the model remains vulnerable to bayesian poisoning     \\ 
\hline

Patricia Iglesias et al. \cite{iglesias2023detecting}   & Coco and ILSVR dataset  & proposal of deep learning based CNN model for detecting drive-by road attack & Convolutional Neural Network & limited to drive-by road attack detection in stego images  \\
\hline

mahdi bahaghighat (2023) \cite{bahaghighat2023high}  & Grega Vrbančič phishing dataset  & performance comparison of phishing detection method based on several six different algorithm  & naive bayes \newline KNN \newline SVM \newline Random Forest \newline Gradient Boost \newline Logistic Regression & 1. complexity of re-generating tree for every output in random forest remains \newline 2. sole reliance on URL feature attributes means the model remains vulnerable to friendly URL   \\ 
\hline

Nishitha U et al (2023) \cite{nishitha2023phishing}  & Kaggle Phishing Dataset  & performance comparison of machine learning and deep learning based algorithm for phishing detection  & CNN \newline RNN \newline KNN \newline Random Forest \newline Decision Tree \newline Logistic Regression & 1. 5000 records is too small to train a CNN model and so no confidence here \newline 2. imbalance in the dataset will lead to bias \newline sole reliance on URL feature   \\ 
\hline

Prince Roy et al. (2022) \cite{sultan2022man}  & Kaggle SQL Injection attack dataset & evaluation of machine learning classifiers for the detection of SQL Injection attacks  & Logistic Regression  \newline Naive Bayes \newline XGBoost \newline Random Forest \newline Adaboost & 1. Limited only to SQL injection, does not address other web application vulnerabilities  \\
\hline

Twana Mustafa and Murat Karabatak (2023) \cite{mustafa2023feature}  & UCL Phishing Dataset & Performance Comparison of different Bayesian classifier based on different Feature Selection Algorithm & naive bayes \newline individual FS \newline forward FS \newline backward FS \newline Plus-I takeaway-r FS \newpage AR1 FS  & 1. each of the Bayesian model remains vulnerable to Bayesian poisoning \newline no hint on why Plus-I takeaway-r FS works better best for naive bayes     \\ 
\hline

Morufu Olalere   & dataset was internally generated from dmoz  & proposal of Naive Bayes model for detecting SQL injection attack & Naive Bayes & SQL syntax distribution in the dataset was not known, It is difficult to know how the proposed model will perform on Kaggle SQL injection dataset \\
\hline

 Jaya T et al (2023) \cite{jaya2023appropriate}  & UK-2011, SpamBase, and Spam Assassin datasets  & usage of frequency weightage of the words for unsupervised clustering of mail into spam and ham messages & naive bayes \newline random forest \newline logistic regression \newline random tree \newline LTSM  & 1. while random forest performed well, its complexity of generating tree for every output remains a problem \newline possible reason for the poor performance of Bayesian classifier remains to be investigated     \\ 
\hline

Santhosh Raminedi et al (2023) \cite{raminedi2023classification}  & Internally collected phishing dataset  & evaluation of several machine learning and deep learning based algorithms for phishing detection using URL features  & ANN \newline SVM \newline KNN and Naive Bayes \newline Random Forest \newline Decision Tree \newline Logistic Regression & 1. complexity of generating tree for every output in random forest was not addressed \newline 2. sole reliance on URL feature   \\ 
\hline

 \end{tabular}
\end{table*}

\section{Background Study}
Technological innovation has led to different methods of securing a system, and so, an appropriate attack technique had to be applied for an attack to be successful. It is for this reason that experienced attacker takes time to study a system to understand it to determine the right approach of a successful attack on a targeted system. Hence, it is imperative to summarize attack methodologies for each of the current state-of-the-art attack techniques as it keeps evolving to understand why some state-of-the-art defense methods remain vulnerable. In this section, we summarized techniques and how some of the major attacks are being perpetrated by cybercriminals.

\subsection{Category of Current State of the art Phishing Detection model}
\subsubsection{Bayesian-Based-Classifier}
Naive Bayes is a family of probabilistic-based algorithms that is based on the Bayes rule. It is based on the fact that, if B has occurred, we can find the probability that A will occur. B is taken to be the evidence while the hypothesis is A and with a strong assumption that each of the features is independent. It uses the prior probability distribution to predict the posterior probability of a sample that belongs to a class. In this process, the class with the highest probability is then selected as the final predicted class \cite{wang2023combination}. Naive Bayes updates prior belief of an event occurring given that there is new information. Hence, given the availability of new data, the probability of the selected sample occurring is given by;
        
        \begin{align*}
            P(class/features) = \frac{P(class) * P(features/class)}{P(features)}
        \end{align*}

\hspace{0.1cm}Where
        
        \begin{itemize}
            \item P(class/features) : Posterior Probability
            \item P(class) : Class Prior Probability
            \item P(features/class) : Likelihood
            \item P(features) : Predictor Prior Probability
        \end{itemize}

It has a very strong assumption of independency which affects its performance for classification tasks\cite {ige2023performance} as the strong assumption of independence among features is not always valid in most of the dataset that is used to train the current state-of-the-art model for several classification tasks. The strong assumption of the Naive Bayes classifier is one reason why it usually underperforms when compared with its peers for similar classification tasks. Naive Bayes classifier has different variants with each variant having its own individual assumption which also impacts its performance in addition to the general assumption of independence which is common to all variants of the Naive Bayes classifier, and so each variant is suitable for different classification tasks.

Multinomial Naive Bayes is a variant of Naive Bayes, It assumes multinomial distribution among features of dataset in addition to the general assumption of independency, and so its performance is affected if the actual distribution is not multinomial or partially multinomial. Multinomial Naive Bayes is the suitable variant for natural language processing classification task \cite{ige2022ai} but still underperforms when compared with non-bayesian and deep learning-based classifiers for the same NLP classification task.

Gaussian Naive Bayes is the suitable Bayesian variant for anomaly detection in network intrusion which could be used to detect Distributed Denial of Service (DDOS) attacks \cite{ige2023performance}. It assumes the normal distribution among features in dataset in addition to the general assumption of independence which is common to all variants of Naive Bayes. The probability density of the normal distribution in Gaussian Naive Bayes is such that:

    \begin{equation}
f(x) = \frac{1}{\sigma \sqrt{2\pi} } e^{-\frac{1}{2}\left(\frac{x-\mu}{\sigma}\right)^2}
\end{equation}
 \hspace{0.1cm}Where
        \begin{itemize}
            \item '${\mu}$' is the mean or expectation of the distribution, 
            \item '${\sigma}$' is the standard deviation, and 
            \item '${\sigma^2}$' is the variance. 
        \end{itemize}

Despite being a suitable Naive Bayes variant for anomaly detection, it still underperforms when compared with its suitable peer for detection of Distributed Denial of Service (DDOS) attack as evident in the work done by Rajendran \cite{rajendran2023improved} where Gaussian Naive Bayes have the least accuracy of 78.75\% compared with other non-bayesian based for attack detection classification task.

Bernoulli Naive Bayes assumes Bernoulli distribution in addition to the assumption of independence. Its main feature is that it only accepts binary values such as success or failure, true or false, and yes or no as input while complement Naive Bayes is used for imbalance datasets as no single variant of Naive Bayes can do the task of all the variants. Both the suitability and performance of each variant are determined by their individual assumption in addition to the general assumption of independence which impacts their performance when compared with their suitable peer for the same classification task.

\subsection{Non-Bayesian Based Classifier}
\subsubsection{Decision Tree}
A decision Tree is a Supervised learning technique whose operation is based on a tree-structured classifier, with features in the dataset being represented by an internal node, each decision rule is represented by the branches, while the internal nodes represent the features of a dataset, branches represent the decision rules and each leaf node represents the decision outcome is represented by the leaf node and so does not have further branches. It makes a decision-based graphical representation of all possible solutions to a problem. It uses the Classification and Regression Tree algorithm (CART) \cite{zhu2020dtof} to construct a decision tree starting with the root node whose branch keeps expanding further to construct a tree-like structure. It is a non-parametric and the ultimate goal is the creation of a machine learning model capable of making prediction by learning simple decision rules that are inferred from data features.

\subsubsection{Random Forest}
It is an ensemble-based learning algorithm that could be used for classification, regression task, and other similar tasks that operates based on the construction of multiple decision trees \cite{hr2020development}. Since the algorithm works by constructing multiple decision trees during training, the output of a classification model trained with a random forest algorithm is the class selected by most of the trees, while the mean or average prediction of individual trees is returned as the output for a regression task. This system of aggregating and ensemblement with multiple trees for prediction makes it possible for a random forest-trained model to outperform the decision tree-trained model and also avoid overfitting which is a peculiar problem for decision tree classifiers.

\subsubsection{Logistic Regression}
Logistic regression is the modeling of the probability of a discrete outcome by having the event log-odds be a linear combination of one or more independent variables given an input variable \cite{EDGAR201795}. Logit transformation is applied to the bounded odds which is the division between the probability of success and probability of failure. It is a linear regression that could be used for both classification and regression tasks and since the output is a probability, the dependent variable is bounded between 0 and 1 values, it uses logistic function to model binary output for classification problems. The difference between linear regression and logistic regression is that the range in logistic regression is bounded by 0 and 1, and also that logistic regression does not require a linear relationship between input and output.

\subsubsection{XGBoost}
It is a supervised learning algorithm that is gradient boosting based. It is extremely efficient and highly scalable, the algorithm works by first creating a series of individual machine learning models and then combining each of the previously created models to form an overall model that is more accurate and efficient than any of the previously created individual models in the series. This system of creating a series of models and combining them to create a single model \cite{gu2022ensemble} makes XGBoost perform better than other state-of-the-art machine learning algorithms in many classification, ranking, several user-defined prediction problems, and regression tasks across several domains. XGboost uses gradient descent to add additional individual models to the main model for prediction, hence it is also known as stochastic gradient boosting, gradient boosting machines, or multiple additive regression trees.

\subsubsection{K-Nearest Neighbor (KNN)}

k-nearest neighbors (kNN) algorithm is a non-parametric supervised learning algorithm that uses the principle of similarity to predict the label or value of a new data point by considering values of its K-nearest neighbors in the training dataset based on a distance metric like  Euclidean distance. 
\begin{equation}
\textrm{dist}(x,z) \leq \textrm{dist}(x,y) + \textrm{dist}(y,z)
\end{equation}

for which the distance between x and z could be calculated by

\begin{equation}
 d\left( x,z\right)   = \sqrt {\sum _{i=1}^{n}  \left( x_{i}-z_{i}\right)^2 } 
\end{equation}

The prediction of the new data point is based on the average or majority vote of its neighbor, this method allows the classifier to adapt its prediction according to the local structure of the data which ultimately helps to improve its overall accuracy and flexibility. Since KNN can be used for both classification and regression tasks, its prediction output depends on the type of task (classification or regression). In the case of a classification task, it uses class membership as the output by using the plurality vote of its neighbor to assign the input to the class that is most common among its k nearest neighbors, but when KNN is being used for a regression task, it uses the average of the values of k nearest neighbors as the prediction output, the value of k has an impact on the overall accuracy \cite{assegie2021k} of the model.

\subsubsection{Support Vector Machine (SVM)}
Support Vector Machine (SVM) is a supervised machine algorithm that works by looking for a hyper-plane that creates a boundary between two classes of data to solve classification and regression-related problems \cite{guo2019improving}. It uses the hyper-plane to determine the best decision boundary between different categories in the training dataset, hence they can be applied to vectors that could encode data. Two theories must hold before we can determine the suitability of SVM for certain classification or regression tasks, the first is the availability of high-dimension input space as SVM tries to prevent overfitting by using an overfitting protective measure which is independent of the number of features in the data gives SVM the potential to handle feature spaces in the dataset. The second theory is the presence of linearly separable properties of categorization in the training dataset, and this is because SVM works by finding linear separators between each of the categories to make accurate predictions.

\subsection{Deep Learning Based Classifier}
\subsubsection{Convolutional Neural Network (CNN)}
CNN is a deep learning model with a grid pattern for processing data that is designed to automatically and adaptively learn spatial hierarchies of features, from low- to high-level patterns \cite{yamashita2018convolutional}, \cite{ige2023adversarial}. It is a mathematical construct that is composed of convolution, pooling, and fully connected layers as three types of layers or building blocks responsible for different tasks for predictions. While convolution and pooling layers, perform feature extraction, the fully connected layer, maps the extracted features into the final output usually known as classification. The convolution layer is composed of mathematical operations (convolution) which plays a very crucial role in Convolutional Neural Networks as in a kind of linear operation. The CNN architecture is a combination of several building blocks like convolution layers, pooling layers, and fully connected layers, and so, a typical architecture consists of repetitions of a stack of many convolution layers and a pooling layer, and then followed by one or more fully connected layers. It stored digital images, and pixel values as a two-dimensional (2D) grid which is an array of numbers along with some parameters called the kernel before an optimizable feature extractor is finally applied at each image position. This makes CNNs a highly efficient classifier for image processing classification tasks, since a feature may occur anywhere in the image. extracted features can hierarchically and progressively become more complex as each layer progressively feeds its output to the next layer, the main task is the minimization of differences between output and ground truth by backward propagation and gradient descent which is an optimization algorithm. This process of optimizing parameters like kernels to minimize the difference between outputs and ground truth is called training.

\subsubsection{Recurrent Neural Network (RNN)}
Recurrent Neural Networks (RNNs) is a type of Neural Network in which output from the previous step is fed to the current step as input, It introduce the concept of memory to neural networks through the addition of the dependency between data points. This addition of dependency between data points ensured that RNNs could be trained to remember concepts by able able to learn repeated patterns. The main difference between RNN and the traditional neural network is the concept of memory in RNN which is made possible as a result of the feedback loop in the cell. Here, it is the feedback loop that enables the possibility of passing information within a layer unlike in feedforward neural networks where information can only be passed between layers. While input and output are independent of each other in a traditional neural network, It is a different ball game in RNN where sequence information is to be remembered, this was made possible in RNN by its Hidden state also known as the memory state through which it remembers previous input to the network, and so it is safe to conclude that the most important features of RNNs is the Hidden state by which it remembers some information in a sequence.
In terms of architecture, RNN architecture is the same as that of other deep neural networks, the main difference lies in how the information flows from the input to the output. While the weight across the network in RNN is the same, deep neural network has different weight matrices for each dense network. The Hidden state in the RNNs which enables them to remember sequence information makes it suitable for natural language processing tasks.

\subsubsection{Long Short-Term Memory (LSTM)}
Long short-term memory (LSTM) network is a recurrent neural network (RNN) that is specifically designed to handle sequential data, such as speech, text, and time series, it is aimed at solving the problem of vanishing gradient in traditional RNNs. It is insensitive to gap length which gives it an advantage over hidden Markov models, hidden Markov models, and other RNNs. It provides a short-term memory for RNN which can last thousands of timesteps thereby making it a "long short-term memory" network. A single LSTM network unit is composed of an output gate, a cell, an input gate, and a forget gate. While the three gates regulate the flow of information into and out of the cell, the cell is responsible for remembering values over arbitrary time intervals as the Forget gates decide on the information to discard from a previous state by assigning a previous state, compared to a current input which assigns a value between 0 and 1. A value of 1 means the information is to be kept, and a value of 0 means the information is to be discarded. The Input gates decide on the exact pieces of new information to store in the current state in the same way as forget gates. Output gates consider both the previous and current states to control which pieces of information in the current state are to output by assigning a value from 0 to 1 to the information. This selective outputting of relevant information from the current state allows the LSTM network to utilize both useful and long-term dependencies in making more accurate predictions in current and future time steps. The fact that they are designed to learn long-term dependencies in sequential data makes them suitable for time series forecasting, speech recognition, and language translation tasks.

\section{Literature Review}

Several methods have been proposed for each of the major categories of cyberattack ranging from Malware, phishing, Man In The Middle, SQL Injection, and Drive-by attack detection and to mitigate the effect of each category of cyberattack with different results and efficiency. These methods are classified based on the different methodologies of the algorithm which can be classified as Bayesian-based, non-Bayesian-based, and deep learning-based. Each of these categories of classification has different accuracy and efficiency for phishing detection and prevention tasks with several underlying causes. In this section, we reviewed and explained existing state-of-the-art phishing detection techniques to identify weak spots where improvement is needed to increase efficiency and project future research direction.

\begin{table}[h]
\centering
\caption{Comparison Analysis of Methodologies for Detection of SQL Injection Attack}
\begin{tabular}{|p{1in}|p{1.5in}|p{0.6in}|}
 \hline
 Classifier & Authors & Mean Score \\
 \hline

 Naive Bayes & \cite{roy2022sql}, \cite{xie2019sql}, \cite{deriba2022development}, \cite{krishnan2021sql}, \cite{pattewar2019detection}, \cite{sivasangari2021sql}, \cite{al2023comparison}  & 90.4  \\
\hline

 SVM & \cite{xie2019sql}, \cite{deriba2022development}, \cite{hasan2019detection}, \cite{krishnan2021sql}, \cite{sivasangari2021sql}, \cite{hosam2021sql},\cite{al2023comparison} & 87.63   \\
\hline

 Random Forest & \cite{roy2022sql}, \cite{xie2019sql}, \cite{sivasangari2021sql}, \cite{hosam2021sql}, \cite{tripathy2020detecting}, \cite{alam2021scamm}, \cite{al2023comparison} &  93.71 \\
\hline

 Logistic Regression & \cite{roy2022sql}, \cite{krishnan2021sql}, \cite{hosam2021sql}, \cite{arumugam2019prediction}, \cite{alam2021scamm}, \cite{al2023comparison}, \cite{bhardwaj2022detection} & 89.68  \\
\hline

KNN & \cite{sivasangari2021sql}, \cite{alam2021scamm}, \cite{al2023comparison}, \cite{adebiyi2021sql}, \cite{hashem2021proposed}, \cite{bhardwaj2022detection}, \cite{irungu2023artificial} &  87.2 \\
\hline

Decision Tree & \cite{hosam2021sql}, \cite{deriba2022development}, \cite{tripathy2020detecting}, \cite{al2023comparison}, \cite{adebiyi2021sql}, \cite{hashem2021proposed}, \cite{ingre2018decision} & 90.04  \\
\hline

\end{tabular}
\label{Comparison Analysis of Methodologies for Detection of SQL Injection Attack}
\end{table}

\begin{table}[h]
\centering
\caption{Comparison Analysis of Methodologies for Detection of DDoS Attack}
\begin{tabular}{|p{1in}|p{1.5in}|p{0.6in}|}
 \hline
 Classifier & Authors & Mean Score \\
 \hline

  SVM & \cite{suresh2011evaluating}, \cite{kyaw2020machine}, \cite{he2017machine}, \cite{alzahrani2021security}, \cite{pei2019ddos} &  90.0   \\
\hline

 Naive Bayes & \cite{suresh2011evaluating}, \cite{saini2020detection}, \cite{he2017machine}, \cite{ajeetha2019machine}, \cite{robinson2015ranking}, \cite{zekri2017ddos} & 84.6   \\
\hline

 Random Forest & \cite{al2023man}, \cite{saini2020detection}, \cite{he2017machine}, \cite{gaur2022analysis}, \cite{ajeetha2019machine}, \cite{robinson2015ranking} &  93.34  \\
\hline

 Decision Tress & \cite{al2023man}, \cite{suresh2011evaluating}, \cite{he2017machine}, \cite{gaur2022analysis}, \cite{alzahrani2021security}, \cite{tuan2020performance} &  96  \\
\hline

 XGBoost & \cite{al2023man}, \cite{gaur2022analysis}, \cite{mohmand2022machine}, \cite{chen2018xgboost}, \cite{rozam2023xgboost} &  96.2  \\
\hline

 KNN & \cite{suresh2011evaluating}, \cite{gaur2022analysis}, \cite{alzahrani2021security}, \cite{doshi2018machine}, \cite{yusof2016evaluation} & 96.5  \\
\hline

\end{tabular}
\label{Comparison Analysis of Methodologies for Detection of DDoS Attack}
\end{table}

\begin{table}[h]
\centering
\caption{Comparison Analysis of Methodologies for Detection of Phishing Attack}
\begin{tabular}{|p{1in}|p{1.5in}|p{0.6in}|}
 \hline
 Classifier & Authors & Mean Score \\
 \hline

 Naive Bayes  & \cite{bahaghighat2023high}, \cite{raminedi2023classification}, \cite{yaswanth2023prediction}, \cite{karim2023phishing}, \cite{Omari2023}, \cite{al2023ofmcdm}  &  80.431  \\
 \hline

  SVM & \cite{bahaghighat2023high}, \cite{raminedi2023classification}, \cite{karim2023phishing}, \cite{Omari2023}, \cite{alnemari2023detecting}, \cite{al2023ofmcdm} & 
   89.429  \\
 \hline

  Random Forest & \cite{bahaghighat2023high}, \cite{raminedi2023classification}, \cite{yaswanth2023prediction}, \cite{karim2023phishing}, \cite{Omari2023}, \cite{almseidin2019phishing} &    97.065  \\
 \hline

  Decision Tree & \cite{raminedi2023classification}, \cite{karim2023phishing}, \cite{Omari2023}, \cite{alnemari2023detecting}, \cite{al2023ofmcdm}, \cite{rashid2023enhanced} &    95.248  \\
 \hline

  Logistic Regression & \cite{bahaghighat2023high}, \cite{raminedi2023classification}, \cite{Omari2023}, \cite{arivukarasi2023efficient}, \cite{pandey2023phish}, \cite{sunday2023phishing} &    92.589   \\
 \hline

  KNN & \cite{bahaghighat2023high}, \cite{raminedi2023classification}, \cite{karim2023phishing}, \cite{Omari2023}, \cite{al2023ofmcdm}, \cite{aldakheel2023deep}     &     90.479  \\
 \hline

\end{tabular}
\label{Comparison Analysis of Methodologies for Detection of Phishing Attack}
\end{table}

\begin{table}[h]
\centering
\caption{Comparison Analysis of Methodologies for Malware Detection}
\begin{tabular}{|p{1in}|p{1.5in}|p{0.6in}|}
 \hline
 Classifier & Authors & Mean Score \\
 \hline

  SVM & \cite{narayanan2016performance}, \cite{udayakumar2018malware}, \cite{mahajan2019malware}, \cite{firdausi2010analysis}, \cite{khammas2015feature}, \cite{shhadat2020use} &  85.5   \\
\hline

 Naive Bayes & \cite{mahajan2019malware}, \cite{masum2022ransomware},\cite{schultz2000data}, \cite{liu2017automatic}, \cite{shabtai2012andromaly}, \cite{firdausi2010analysis}  & 79.6   \\
\hline

 Random Forest & \cite{mahajan2019malware}, \cite{pirscoveanu2015analysis}, \cite{chumachenko2017machine}, \cite{sethi2018novel}, \cite{agarkar2020malware}, \cite{masum2022ransomware} &  96.04  \\
\hline

 Decision Tree & \cite{mahajan2019malware}, \cite{sethi2018novel}, \cite{agarkar2020malware}, \cite{masum2022ransomware}, \cite{liu2017automatic}, \cite{shhadat2020use}  &  91.6  \\
\hline

 KNN & \cite{narayanan2016performance}, \cite{liu2017automatic}, \cite{mahajan2019malware}, \cite{shabtai2012andromaly}, \cite{firdausi2010analysis}, \cite{shhadat2020use}  & 86.6  \\
\hline

Logistic Regression & \cite{liu2017automatic}, \cite{shabtai2012andromaly}, \cite{bae2020ransomware}, \cite{suhuan2019android}, \cite{masum2022ransomware}, \cite{kumar2017logistic}  & 90.06  \\
\hline

\end{tabular}
\label{Comparison Analysis of Methodologies for Malware Detection}
\end{table}

\begin{table}[h]
\centering
\caption{Comparison Analysis of Methodologies for Man-in-the-Middle Attack Detection}
\begin{tabular}{|p{1in}|p{1.5in}|p{0.6in}|}
 \hline
 Classifier & Authors & Mean Score \\
 \hline

  SVM & \cite{das2022man}, \cite{manhas2021implementation}, \cite{alsulami2022intrusion}, \cite{ham2014linear}, \cite{ioannou2019classifying}, \cite{hasan2019attack} &  92.3   \\
\hline

 Naive Bayes & \cite{manhas2021implementation}, \cite{al2020using}, \cite{vijayalakshmi2023detection}, \cite{koppula2023lnkdsea}, \cite{ussatova2024development}, \cite{priya2022network} & 82.8   \\
\hline

 Random Forest & \cite{das2022man}, \cite{al2020using}, \cite{vijayalakshmi2023detection}, \cite{hasan2019attack}, \cite{desai2020iot}, \cite{reji2023anomaly} &  96.6  \\
\hline

 Decision Tree & \cite{das2022man}, \cite{manhas2021implementation}, \cite{alsulami2022intrusion}, \cite{hasan2019attack}, \cite{desai2020iot}, \cite{koppula2023lnkdsea}  &  88.6  \\
\hline

 KNN & \cite{das2022man}, \cite{manhas2021implementation}, \cite{al2020using}, \cite{alsulami2022intrusion}, \cite{vijayalakshmi2023detection}, \cite{koppula2023lnkdsea} & 91.5  \\
\hline

Logistic Regression & \cite{das2022man}, \cite{hasan2019attack}, \cite{koppula2023lnkdsea}, \cite{priya2022network}, \cite{toutsop2020monitoring}, \cite{chalichalamala2023logistic} & 84.8  \\
\hline

\end{tabular}
\label{Comparison Analysis of Methodologies for Man-in-the-Middle Attack Detection}
\end{table}

\begin{table}[h]
\centering
\caption{Comparison Analysis of Methodologies for Drive-By Attack Detection}
\begin{tabular}{|p{1in}|p{1.5in}|p{0.6in}|}
 \hline
 Classifier & Authors & Mean Score \\
 \hline

  SVM & \cite{adachi2015approach}, \cite{priya2013static}, \cite{cherukuri2014detection}, \cite{aljabri2022phishing}, \cite{raja2022malicious}, \cite{jain2022apuml} &  88.69   \\
\hline

 Naive Bayes & \cite{javed2017real}, \cite{adachi2015approach}, \cite{ibrahim2019detection}, \cite{thao2017classification}, \cite{pandeydetection}, \cite{sahu2021malignant} & 87.37   \\
\hline

 Random Forest & \cite{adachi2015approach}, \cite{aljabri2022phishing}, \cite{raja2022malicious}, \cite{jain2022apuml}, \cite{pandeydetection}, \cite{catak2021malicious} &  91.83  \\
\hline

 Decision Tree & \cite{adachi2015approach}, \cite{harnmetta2018classification}, \cite{kikuchi2015automated}, \cite{jain2022apuml}, \cite{thao2017classification}, \cite{eshete2013binspect} &  91.85  \\
\hline

 KNN & \cite{priya2013static}, \cite{jain2022apuml}, \cite{thao2017classification}, \cite{pandeydetection}, \cite{sahu2021malignant}, \cite{okpanachi2019machine} & 92.22  \\
\hline

Logistic Regression & \cite{priya2013static}, \cite{jain2022apuml}, \cite{pandeydetection}, \cite{eshete2013binspect}, \cite{chiramdasu2021malicious}, \cite{vanitha2019malicious} & 92.76  \\
\hline

\end{tabular}
\label{Comparison Analysis of Methodologies for Drive-By Attack Detection}
\end{table}

\section{Analysis and Discussion}

\begin{figure*}
\centering
\includegraphics[width=0.8\linewidth]{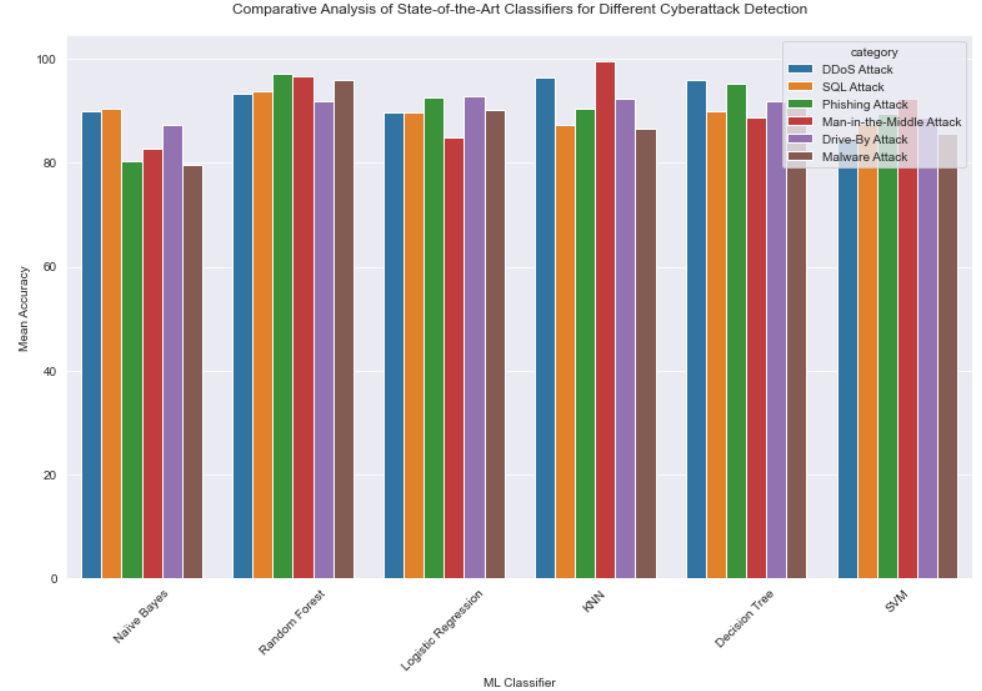}
\caption{\label{fig: Cyber Security Attacks } Comparative Analysis of Algorithms for Detection of Different Cyberattacks}
\end{figure*}

1. Insufficient research on the capability of machine learning algorithms for the detection of drive-by downloads, man-in-the-middle, and Malware attacks. \newline 

We got very few research papers where a machine learning algorithm was used to train a model for the detection of drive-by downloads, man-in-the-middle, and Malware attacks or any of their combination. The few research papers where machine learning algorithms were used to train models for the detection of drive-by download or man-in-the-middle were so few that comparing the result with the performance of machine learning algorithms for the detection of other categories of cyberattack will lead to severe bias on the result by tilting it against the performance of ML models in the detection of other categories of cyberattacks. Research papers where machine learning algorithms were used to detect drive-by download attacks were much more scanty than the other two. Hence, we chose to remove them from the relative comparison table as more research work where various machine learning algorithms are used to detect drive-by download attack are still required. We don't know why the detection of drive-by download attacks using machine learning algorithms is so scanty, and so this is open for investigation and further research. \newline

2. Mix performance by Naive Bayes \newline 

An observation of the performance of Naive Bayes across different categories of cyberattack is very interesting knowing fully well that Naive Bayes is a parametric-based machine learning algorithm whose prediction is based on (i)the assumption of independence between features and (ii) the distribution of features in a dataset which might be Multinomial, Bernoulli, or Normal distribution. We expect Naive Bayes to have relatively consistent performance across different categories of cyberattacks based on the assumption but it was surprising to see Naive Bayes having a very strong performance in the detection of SQL Injection attacks while having the weakest performance in the detection of phishing attacks in relative comparison to other machine learning classifier. Understanding why Naive Bayes algorithm performs extremely well in the detection of SQL injection attacks but relatively poor performance in the detection of phishing attacks requires further research. \newline 

3. Limitation of Current Approach to SQL Injection Attack Detection \newline 

The current approach to the detection of Structured Query Language Injection (SQLI) attacks is solely based on the presence of SQL statements in user input, and while this approach had been successful in finding the presence of SQL statements in user input to protect backend database against SQL injection attack, the approach cannot detect an already compromised database. Naive Bayes and Random Forest are the best-performing machine learning algorithms with mean accuracies of 90.4\% and 93.71\% respectively for the detection of SQLi attacks based on the current approach of finding the presence of SQL statements from user input. Hence, with a failure of 9.6\% for Naive Bayes and 6.3\% for Randon Forest algorithms, a database can still be compromised, hence machine learning model needs to have the capability to predict a compromised database immediately an SQLi attack scale through. There is no single study on the detection of compromise of a database by a machine learning model, hence, this is an interesting area that requires further research.

\section{Conclusion and Research Direction}

In this research, we did a comprehensive survey of current state-of-the-art machine learning algorithms to investigate their effectiveness and suitability for the detection of different categories of cyberattacks, to ensure that our research reflects the latest advancement at the intersection of artificial intelligence and cybersecurity, we categorized and discussed various methodologies, techniques, and approaches from research papers that are from the past 10 years but predominantly from the last 5 years. We also reviewed the effectiveness and limitations of recent proposals and novel frameworks in the detection of cyberattacks. Our finding shows the need for; further research and exploration on the use of a machine learning approach for the detection of drive-by download attacks, an investigation into the mix performance of Naive Bayes to identify possible research direction on improvement to existing state-of-the-art Naive Bayes classifier, we also identify the need for an improvement to the current machine learning approach to the detection of SQLi attack because existing machine learning approach cannot detect an already compromised database with SQLi attack.

\bibliographystyle{plain}
\bibliography{main.bib}

\begin{thebibliography}{100}

\bibitem{adachi2015approach}
Takashi Adachi and Kazumasa Omote.
\newblock An approach to predict drive-by-download attacks by vulnerability evaluation and opcode.
\newblock In {\em 2015 10th Asia Joint Conference on Information Security}, pages 145--151. IEEE, 2015.

\bibitem{adebiyi2021sql}
Marion~Olubunmi Adebiyi, Micheal~Olaolu Arowolo, Goodnews~Ime Archibong, Moses~Damilola Mshelia, and Ayodele~Ariyo Adebiyi.
\newblock An sql injection detection model using chi-square with classification techniques.
\newblock In {\em 2021 International Conference on Electrical, Computer and Energy Technologies (ICECET)}, pages 1--8. IEEE, 2021.

\bibitem{agarkar2020malware}
Sanket Agarkar and Soma Ghosh.
\newblock Malware detection \& classification using machine learning.
\newblock In {\em 2020 IEEE International Symposium on Sustainable Energy, Signal Processing and Cyber Security (iSSSC)}, pages 1--6. IEEE, 2020.

\bibitem{agarwal2016machine}
Mayank Agarwal, Dileep Pasumarthi, Santosh Biswas, and Sukumar Nandi.
\newblock Machine learning approach for detection of flooding dos attacks in 802.11 networks and attacker localization.
\newblock {\em International Journal of Machine Learning and Cybernetics}, 7:1035--1051, 2016.

\bibitem{ajeetha2019machine}
G~Ajeetha and G~Madhu Priya.
\newblock Machine learning based ddos attack detection.
\newblock In {\em 2019 Innovations in Power and Advanced Computing Technologies (i-PACT)}, volume~1, pages 1--5. IEEE, 2019.

\bibitem{al2023ofmcdm}
Md~Abdullah Al~Ahasan, Mengjun Hu, and Nashid Shahriar.
\newblock Ofmcdm/irf: A phishing website detection model based on optimized fuzzy multi-criteria decision-making and improved random forest.
\newblock In {\em 2023 Silicon Valley Cybersecurity Conference (SVCC)}, pages 1--8. IEEE, 2023.

\bibitem{al2020using}
Mousa Al-Akhras, Mohammed Alawairdhi, Ali Alkoudari, and Samer Atawneh.
\newblock Using machine learning to build a classification model for iot networks to detect attack signatures.
\newblock {\em Int. J. Comput. Netw. Commun.(IJCNC)}, 12:99--116, 2020.

\bibitem{al2023man}
Sura Abdulmunem~Mohammed Al-Juboori, Firas Hazzaa, Zinah~Sattar Jabbar, Sinan Salih, and Hassan~Muwafaq Gheni.
\newblock Man-in-the-middle and denial of service attacks detection using machine learning algorithms.
\newblock {\em Bulletin of Electrical Engineering and Informatics}, 12(1):418--426, 2023.

\bibitem{al2023comparison}
Manar Hasan~Ali AL-Maliki and Mahdi~Nsaif Jasim.
\newblock Comparison study for nlp using machine learning techniques to detecting sql injection vulnerabilities.
\newblock {\em International Journal of Nonlinear Analysis and Applications}, 2023.

\bibitem{alam2021scamm}
Auninda Alam, Marjan Tahreen, Md~Moin Alam, Shahnewaz~Ali Mohammad, and Shohag Rana.
\newblock {\em SCAMM: Detection and prevention of SQL injection attacks using a machine learning approach}.
\newblock PhD thesis, Brac University, 2021.

\bibitem{aldakheel2023deep}
Eman~Abdullah Aldakheel, Mohammed Zakariah, Ghada~Abdalaziz Gashgari, Fahdah~A Almarshad, and Abdullah~IA Alzahrani.
\newblock A deep learning-based innovative technique for phishing detection in modern security with uniform resource locators.
\newblock {\em Sensors}, 23(9):4403, 2023.

\bibitem{aljabri2022phishing}
Malak Aljabri and Samiha Mirza.
\newblock Phishing attacks detection using machine learning and deep learning models.
\newblock In {\em 2022 7th International Conference on Data Science and Machine Learning Applications (CDMA)}, pages 175--180. IEEE, 2022.

\bibitem{almseidin2019phishing}
Mohammad Almseidin, AlMaha~Abu Zuraiq, Mouhammd Al-Kasassbeh, and Nidal Alnidami.
\newblock Phishing detection based on machine learning and feature selection methods.
\newblock 2019.

\bibitem{alnemari2023detecting}
Shouq Alnemari and Majid Alshammari.
\newblock Detecting phishing domains using machine learning.
\newblock {\em Applied Sciences}, 13(8):4649, 2023.

\bibitem{alsulami2022intrusion}
Abdulaziz~A Alsulami, Qasem Abu Al-Haija, Ahmad Tayeb, and Ali Alqahtani.
\newblock An intrusion detection and classification system for iot traffic with improved data engineering.
\newblock {\em Applied Sciences}, 12(23):12336, 2022.

\bibitem{alzahrani2021security}
Rami~J Alzahrani and Ahmed Alzahrani.
\newblock Security analysis of ddos attacks using machine learning algorithms in networks traffic.
\newblock {\em Electronics}, 10(23):2919, 2021.

\bibitem{arivukarasi2023efficient}
M~Arivukarasi, A~Manju, R~Kaladevi, Shanmugasundaram Hariharan, M~Mahasree, and Andraju~Bhanu Prasad.
\newblock Efficient phishing detection and prevention using support vector machine (svm) algorithm.
\newblock In {\em 2023 IEEE 12th International Conference on Communication Systems and Network Technologies (CSNT)}, pages 545--548. IEEE, 2023.

\bibitem{arumugam2019prediction}
Chamundeswari Arumugam, Varsha~Bhargavi Dwarakanathan, S~Gnanamary, Vishalraj~Natarajan Neyveli, Rohit~Kanakuppaliyalil Ramesh, Yeshwanthraa Kandhavel, and Sadhanandhan Balakrishnan.
\newblock Prediction of sql injection attacks in web applications.
\newblock In {\em Computational Science and Its Applications--ICCSA 2019: 19th International Conference, Saint Petersburg, Russia, July 1--4, 2019, Proceedings, Part IV 19}, pages 496--505. Springer, 2019.

\bibitem{assegie2021k}
Tsehay~Admassu Assegie.
\newblock K-nearest neighbor based url identification model for phishing attack detection.
\newblock {\em Indian Journal of Artificial Intelligence and Neural Networking}, 1:18--21, 2021.

\bibitem{bae2020ransomware}
Seong~Il Bae, Gyu~Bin Lee, and Eul~Gyu Im.
\newblock Ransomware detection using machine learning algorithms.
\newblock {\em Concurrency and Computation: Practice and Experience}, 32(18):e5422, 2020.

\bibitem{bahaghighat2023high}
Mahdi Bahaghighat, Majid Ghasemi, and Figen Ozen.
\newblock A high-accuracy phishing website detection method based on machine learning.
\newblock {\em Journal of Information Security and Applications}, 77:103553, 2023.

\bibitem{bhardwaj2022detection}
Aashutosh Bhardwaj, Saheb~Singh Chandok, Aniket Bagnawar, Shubham Mishra, and Deepak Uplaonkar.
\newblock Detection of cyber attacks: Xss, sqli, phishing attacks and detecting intrusion using machine learning algorithms.
\newblock In {\em 2022 IEEE Global Conference on Computing, Power and Communication Technologies (GlobConPT)}, pages 1--6. IEEE, 2022.

\bibitem{catak2021malicious}
Ferhat~Ozgur Catak, Kevser Sahinbas, and Volkan D{\"o}rtkarde{\c{s}}.
\newblock Malicious url detection using machine learning.
\newblock In {\em Artificial intelligence paradigms for smart cyber-physical systems}, pages 160--180. IGI global, 2021.

\bibitem{chalichalamala2023logistic}
Silpa Chalichalamala, Niranjana Govindan, and Ramani Kasarapu.
\newblock Logistic regression ensemble classifier for intrusion detection system in internet of things.
\newblock {\em Sensors}, 23(23):9583, 2023.

\bibitem{chen2018xgboost}
Zhuo Chen, Fu~Jiang, Yijun Cheng, Xin Gu, Weirong Liu, and Jun Peng.
\newblock Xgboost classifier for ddos attack detection and analysis in sdn-based cloud.
\newblock In {\em 2018 IEEE international conference on big data and smart computing (bigcomp)}, pages 251--256. IEEE, 2018.

\bibitem{cherukuri2014detection}
Manoj Cherukuri, Srinivas Mukkamala, and Dongwan Shin.
\newblock Detection of plugin misuse drive-by download attacks using kernel machines.
\newblock In {\em 10th IEEE International Conference on Collaborative Computing: Networking, Applications and Worksharing}, pages 546--553. IEEE, 2014.

\bibitem{chiramdasu2021malicious}
Rupa Chiramdasu, Gautam Srivastava, Sweta Bhattacharya, Praveen~Kumar Reddy, and Thippa~Reddy Gadekallu.
\newblock Malicious url detection using logistic regression.
\newblock In {\em 2021 IEEE International Conference on Omni-Layer Intelligent Systems (COINS)}, pages 1--6. IEEE, 2021.

\bibitem{chumachenko2017machine}
Kateryna Chumachenko.
\newblock Machine learning methods for malware detection and classification.
\newblock 2017.

\bibitem{das2022man}
Krittika Das, Rajdeep Basu, and Raja Karmakar.
\newblock Man-in-the-middle attack detection using ensemble learning.
\newblock In {\em 2022 13th International Conference on Computing Communication and Networking Technologies (ICCCNT)}, pages 1--6. IEEE, 2022.

\bibitem{deriba2022development}
F~Deriba, Ayodeji~Olalekan Salau, Shaimaa~Hadi Mohammed, Tsegay~Mullu Kassa, and Wubetu~Barud Demilie.
\newblock Development of a compressive framework using machine learning approaches for sql injection attacks.
\newblock 1(7):183--189, 2022.

\bibitem{desai2020iot}
Madhuri~Gurunathrao Desai, Yong Shi, and Kun Suo.
\newblock Iot bonet and network intrusion detection using dimensionality reduction and supervised machine learning.
\newblock In {\em 2020 11th IEEE Annual Ubiquitous Computing, Electronics \& Mobile Communication Conference (UEMCON)}, pages 0316--0322. IEEE, 2020.

\bibitem{doshi2018machine}
Rohan Doshi, Noah Apthorpe, and Nick Feamster.
\newblock Machine learning ddos detection for consumer internet of things devices.
\newblock In {\em 2018 IEEE Security and Privacy Workshops (SPW)}, pages 29--35. IEEE, 2018.

\bibitem{EDGAR201795}
Thomas~W. Edgar and David~O. Manz.
\newblock Chapter 4 - exploratory study.
\newblock In Thomas~W. Edgar and David~O. Manz, editors, {\em Research Methods for Cyber Security}, pages 95--130. Syngress, 2017.

\bibitem{eshete2013binspect}
Birhanu Eshete, Adolfo Villafiorita, and Komminist Weldemariam.
\newblock Binspect: Holistic analysis and detection of malicious web pages.
\newblock In {\em Security and Privacy in Communication Networks: 8th International ICST Conference, SecureComm 2012, Padua, Italy, September 3-5, 2012. Revised Selected Papers 8}, pages 149--166. Springer, 2013.

\bibitem{firdausi2010analysis}
Ivan Firdausi, Alva Erwin, Anto~Satriyo Nugroho, et~al.
\newblock Analysis of machine learning techniques used in behavior-based malware detection.
\newblock In {\em 2010 second international conference on advances in computing, control, and telecommunication technologies}, pages 201--203. IEEE, 2010.

\bibitem{gaur2022analysis}
Vimal Gaur and Rajneesh Kumar.
\newblock Analysis of machine learning classifiers for early detection of ddos attacks on iot devices.
\newblock {\em Arabian Journal for Science and Engineering}, 47(2):1353--1374, 2022.

\bibitem{gu2022ensemble}
Jiaqi Gu and Hui Xu.
\newblock An ensemble method for phishing websites detection based on xgboost.
\newblock In {\em 2022 14th international conference on computer research and development (ICCRD)}, pages 214--219. IEEE, 2022.

\bibitem{guo2019improving}
Bao Guo, Chunxia Zhang, Junmin Liu, and Xiaoyi Ma.
\newblock Improving text classification with weighted word embeddings via a multi-channel textcnn model.
\newblock {\em Neurocomputing}, 363:366--374, 2019.

\bibitem{ham2014linear}
Hyo-Sik Ham, Hwan-Hee Kim, Myung-Sup Kim, Mi-Jung Choi, et~al.
\newblock Linear svm-based android malware detection for reliable iot services.
\newblock {\em Journal of Applied Mathematics}, 2014, 2014.

\bibitem{harnmetta2018classification}
Sukritta Harnmetta and Sudsanguan Ngamsuriyaroj.
\newblock Classification of exploit-kit behaviors via machine learning approach.
\newblock In {\em 2018 20th International Conference on Advanced Communication Technology (ICACT)}, pages 468--473. IEEE, 2018.

\bibitem{hasan2019attack}
Mahmudul Hasan, Md~Milon Islam, Md~Ishrak~Islam Zarif, and MMA Hashem.
\newblock Attack and anomaly detection in iot sensors in iot sites using machine learning approaches.
\newblock {\em Internet of Things}, 7:100059, 2019.

\bibitem{hasan2019detection}
Musaab Hasan, Zayed Balbahaith, and Mohammed Tarique.
\newblock Detection of sql injection attacks: a machine learning approach.
\newblock In {\em 2019 International Conference on Electrical and Computing Technologies and Applications (ICECTA)}, pages 1--6. IEEE, 2019.

\bibitem{hashem2021proposed}
Istiaque Hashem, Minhajul Islam, Shazid~Morshedul Haque, Zobaidul~Islam Jabed, and Nazmus Sakib.
\newblock A proposed technique for simultaneously detecting ddos and sql injection attacks.
\newblock {\em Int. J. Comput. Appl}, 183(11):50--57, 2021.

\bibitem{he2017machine}
Zecheng He, Tianwei Zhang, and Ruby~B Lee.
\newblock Machine learning based ddos attack detection from source side in cloud.
\newblock In {\em 2017 IEEE 4th International Conference on Cyber Security and Cloud Computing (CSCloud)}, pages 114--120. IEEE, 2017.

\bibitem{hosam2021sql}
Eman Hosam, Hagar Hosny, Walaa Ashraf, and Ahmed~S Kaseb.
\newblock Sql injection detection using machine learning techniques.
\newblock In {\em 2021 8th International Conference on Soft Computing \& Machine Intelligence (ISCMI)}, pages 15--20. IEEE, 2021.

\bibitem{hr2020development}
Mohith~Gowda HR, Adithya MV, et~al.
\newblock Development of anti-phishing browser based on random forest and rule of extraction framework.
\newblock {\em Cybersecurity}, 3(1):1--14, 2020.

\bibitem{ibrahim2019detection}
Saeed Ibrahim, Nawwaf~Al Herami, Ebrahim~Al Naqbi, and Monther Aldwairi.
\newblock Detection and analysis of drive-by downloads and malicious websites.
\newblock In {\em International Symposium on Security in Computing and Communication}, pages 72--86. Springer, 2019.

\bibitem{ige2022ai}
Tosin Ige and Sikiru Adewale.
\newblock Ai powered anti-cyber bullying system using machine learning algorithm of multinomial na{\"\i}ve bayes and optimized linear support vector machine.
\newblock {\em arXiv preprint arXiv:2207.11897}, 2022.

\bibitem{ige2023performance}
Tosin Ige and Christopher Kiekintveld.
\newblock Performance comparison and implementation of bayesian variants for network intrusion detection.
\newblock {\em arXiv preprint arXiv:2308.11834}, 2023.

\bibitem{ige2023adversarial}
Tosin Ige, William Marfo, Justin Tonkinson, Sikiru Adewale, and Bolanle~Hafiz Matti.
\newblock Adversarial sampling for fairness testing in deep neural network.
\newblock {\em arXiv preprint arXiv:2303.02874}, 2023.

\bibitem{iglesias2023detecting}
Patricia Iglesias, Miguel-Angel Sicilia, and Elena Garc{\'\i}a-Barriocanal.
\newblock Detecting browser drive-by exploits in images using deep learning.
\newblock {\em Electronics}, 12(3):473, 2023.

\bibitem{ingre2018decision}
Bhupendra Ingre, Anamika Yadav, and Atul~Kumar Soni.
\newblock Decision tree based intrusion detection system for nsl-kdd dataset.
\newblock In {\em Information and Communication Technology for Intelligent Systems (ICTIS 2017)-Volume 2 2}, pages 207--218. Springer, 2018.

\bibitem{ioannou2019classifying}
Christiana Ioannou and Vasos Vassiliou.
\newblock Classifying security attacks in iot networks using supervised learning.
\newblock In {\em 2019 15th International conference on distributed computing in sensor systems (DCOSS)}, pages 652--658. IEEE, 2019.

\bibitem{irungu2023artificial}
John Irungu, Steffi Graham, Anteneh Girma, and Thabet Kacem.
\newblock Artificial intelligence techniques for sql injection attack detection.
\newblock In {\em Proceedings of the 2023 8th International Conference on Intelligent Information Technology}, pages 38--45, 2023.

\bibitem{jain2022apuml}
Ankit~Kumar Jain, Ninmoy Debnath, and Arvind~Kumar Jain.
\newblock Apuml: an efficient approach to detect mobile phishing webpages using machine learning.
\newblock {\em Wireless Personal Communications}, 125(4):3227--3248, 2022.

\bibitem{javed2017real}
Amir Javed, Pete Burnap, and Omer Rana.
\newblock Real time prediction of drive by download attacks on twitter.
\newblock {\em arXiv preprint arXiv:1708.05831}, 2017.

\bibitem{jaya2023appropriate}
T~Jaya, R~Kanyaharini, and Bandi Navaneesh.
\newblock Appropriate detection of ham and spam emails using machine learning algorithm.
\newblock In {\em 2023 International Conference on Advances in Computing, Communication and Applied Informatics (ACCAI)}, pages 1--5. IEEE, 2023.

\bibitem{karim2023phishing}
Abdul Karim, Mobeen Shahroz, Khabib Mustofa, Samir~Brahim Belhaouari, and S~Ramana~Kumar Joga.
\newblock Phishing detection system through hybrid machine learning based on url.
\newblock {\em IEEE Access}, 11:36805--36822, 2023.

\bibitem{khammas2015feature}
Ban~Mohammed Khammas, Alireza Monemi, J~Stephen Bassi, Ismahani Ismail, S~Mohd Nor, and Muhammad~Nadzir Marsono.
\newblock Feature selection and machine learning classification for malware detection.
\newblock {\em Jurnal Teknologi}, 77(1):243--250, 2015.

\bibitem{kikuchi2015automated}
Hiroaki Kikuchi, Hiroaki Matsumoto, and Hiroshi Ishii.
\newblock Automated detection of drive-by download attack.
\newblock In {\em 2015 9th International Conference on Innovative Mobile and Internet Services in Ubiquitous Computing}, pages 511--515. IEEE, 2015.

\bibitem{koppula2023lnkdsea}
Manasa Koppula, Leo~Joseph LM, et~al.
\newblock Lnkdsea: Machine learning based iot/iiot attack detection method.
\newblock In {\em 2023 International Conference on Advances in Electronics, Communication, Computing and Intelligent Information Systems (ICAECIS)}, pages 655--662. IEEE, 2023.

\bibitem{krishnan2021sql}
SS~Anandha Krishnan, Adhil~N Sabu, Priya~P Sajan, and AL~Sreedeep.
\newblock Sql injection detection using machine learning.
\newblock {\em vol}, 11:11, 2021.

\bibitem{kumar2017logistic}
B~Jyothi Kumar, H~Naveen, B~Praveen Kumar, Sai~Shyam Sharma, and Jaime Villegas.
\newblock Logistic regression for polymorphic malware detection using anova f-test.
\newblock In {\em 2017 International Conference on Innovations in Information, Embedded and Communication Systems (ICIIECS)}, pages 1--5. IEEE, 2017.

\bibitem{kyaw2020machine}
Aye~Thandar Kyaw, May~Zin Oo, and Chit~Su Khin.
\newblock Machine-learning based ddos attack classifier in software defined network.
\newblock In {\em 2020 17th International Conference on Electrical Engineering/Electronics, Computer, Telecommunications and Information Technology (ECTI-CON)}, pages 431--434. IEEE, 2020.

\bibitem{liu2017automatic}
Liu Liu, Bao-sheng Wang, Bo~Yu, and Qiu-xi Zhong.
\newblock Automatic malware classification and new malware detection using machine learning.
\newblock {\em Frontiers of Information Technology \& Electronic Engineering}, 18(9):1336--1347, 2017.

\bibitem{magdacy2023detect}
Ann Zeki~Ablahd Magdacy~Jerjes, Adnan~Yousif Dawod, and Mohammed~Fakhrulddin Abdulqader.
\newblock Detect malicious web pages using naive bayesian algorithm to detect cyber threats.
\newblock {\em Wireless Personal Communications}, pages 1--13, 2023.

\bibitem{mahajan2019malware}
Ginika Mahajan, Bhavna Saini, and Shivam Anand.
\newblock Malware classification using machine learning algorithms and tools.
\newblock In {\em 2019 Second international conference on advanced computational and communication paradigms (ICACCP)}, pages 1--8. IEEE, 2019.

\bibitem{manhas2021implementation}
Jatinder Manhas and Shallu Kotwal.
\newblock Implementation of intrusion detection system for internet of things using machine learning techniques.
\newblock {\em Multimedia Security: Algorithm Development, Analysis and Applications}, pages 217--237, 2021.

\bibitem{masum2022ransomware}
Mohammad Masum, Md~Jobair~Hossain Faruk, Hossain Shahriar, Kai Qian, Dan Lo, and Muhaiminul~Islam Adnan.
\newblock Ransomware classification and detection with machine learning algorithms.
\newblock In {\em 2022 IEEE 12th Annual Computing and Communication Workshop and Conference (CCWC)}, pages 0316--0322. IEEE, 2022.

\bibitem{mohmand2022machine}
Muhammad~Ismail Mohmand, Hameed Hussain, Ayaz~Ali Khan, Ubaid Ullah, Muhammad Zakarya, Aftab Ahmed, Mushtaq Raza, Izaz~Ur Rahman, Muhammad Haleem, et~al.
\newblock A machine learning-based classification and prediction technique for ddos attacks.
\newblock {\em IEEE Access}, 10:21443--21454, 2022.

\bibitem{mustafa2023feature}
Twana Mustafa and Murat Karabatak.
\newblock Feature selection for phishing website by using naive bayes classifier.
\newblock In {\em 2023 11th International Symposium on Digital Forensics and Security (ISDFS)}, pages 1--4. IEEE, 2023.

\bibitem{narayanan2016performance}
Barath~Narayanan Narayanan, Ouboti Djaneye-Boundjou, and Temesguen~M Kebede.
\newblock Performance analysis of machine learning and pattern recognition algorithms for malware classification.
\newblock In {\em 2016 IEEE national aerospace and electronics conference (NAECON) and ohio innovation summit (OIS)}, pages 338--342. IEEE, 2016.

\bibitem{nishitha2023phishing}
U~Nishitha, Revanth Kandimalla, Reddy M~Mourya Vardhan, and U~Kumaran.
\newblock Phishing detection using machine learning techniques.
\newblock In {\em 2023 3rd Asian Conference on Innovation in Technology (ASIANCON)}, pages 1--6. IEEE, 2023.

\bibitem{okpanachi2019machine}
Ahiaba~Moses Okpanachi and Morufu Olalere.
\newblock Machine learning approach for detection of spam url: Performance evaluation of selected algorithms.
\newblock 2019.

\bibitem{Omari2023}
Kamal Omari.
\newblock Comparative study of machine learning algorithms for phishing website detection.
\newblock {\em International Journal of Advanced Computer Science and Applications}, 14(9), 2023.

\bibitem{pandey2023phish}
Pankaj Pandey and Nishchol Mishra.
\newblock Phish-sight: a new approach for phishing detection using dominant colors on web pages and machine learning.
\newblock {\em International Journal of Information Security}, pages 1--11, 2023.

\bibitem{pandeydetection}
Shubham Pandey and AK~Malviya.
\newblock Detection of malign pages using machine learning techniques: Description and analysis.

\bibitem{pattewar2019detection}
Tareek Pattewar, Hitesh Patil, Harshada Patil, Neha Patil, Muskan Taneja, and Tushar Wadile.
\newblock Detection of sql injection using machine learning: a survey.
\newblock {\em Int. Res. J. Eng. Technol.(IRJET)}, 6(11):239--246, 2019.

\bibitem{pei2019ddos}
Jiangtao Pei, Yunli Chen, and Wei Ji.
\newblock A ddos attack detection method based on machine learning.
\newblock In {\em Journal of Physics: Conference Series}, volume 1237, page 032040. IOP Publishing, 2019.

\bibitem{pirscoveanu2015analysis}
Radu~S Pirscoveanu, Steven~S Hansen, Thor~MT Larsen, Matija Stevanovic, Jens~Myrup Pedersen, and Alexandre Czech.
\newblock Analysis of malware behavior: Type classification using machine learning.
\newblock In {\em 2015 International conference on cyber situational awareness, data analytics and assessment (CyberSA)}, pages 1--7. IEEE, 2015.

\bibitem{priya2013static}
M~Priya, L~Sandhya, and Ciza Thomas.
\newblock A static approach to detect drive-by-download attacks on webpages.
\newblock In {\em 2013 International Conference on Control Communication and Computing (ICCC)}, pages 298--303. IEEE, 2013.

\bibitem{priya2022network}
N~Shanmuga Priya, S~Meyyappan, KN~Balasubramanian, and AS~Pruthiev.
\newblock Network attack detection using machine learning.
\newblock In {\em 2022 8th International Conference on Advanced Computing and Communication Systems (ICACCS)}, volume~1, pages 342--346. IEEE, 2022.

\bibitem{raja2022malicious}
A~Saleem Raja, B~Sundarvadivazhagan, R~Vijayarangan, and S~Veeramani.
\newblock Malicious webpage classification based on web content features using machine learning and deep learning.
\newblock In {\em 2022 International Conference on Green Energy, Computing and Sustainable Technology (GECOST)}, pages 314--319. IEEE, 2022.

\bibitem{rajendran2023improved}
T~Rajendran, E~Abishekraj, and U~Dhanush.
\newblock Improved intrusion detection system that uses machine learning techniques to proactively defend ddos attack.
\newblock In {\em ITM Web of Conferences}, volume~56, page 05011. EDP Sciences, 2023.

\bibitem{raminedi2023classification}
Santhosh Raminedi, Trilok~Nath Pandey, Venkat~Amith Woonna, Sletzer~Concy Mascarenhas, and Arjun Bharani.
\newblock Classification of phishing websites using machine learning models.
\newblock In {\em 2023 3rd International conference on Artificial Intelligence and Signal Processing (AISP)}, pages 1--5. IEEE, 2023.

\bibitem{rashid2023enhanced}
Saba~Hussein Rashid and Wisam~Dawood Abdullah.
\newblock Enhanced website phishing detection based on the cyber kill chain and cloud computing.
\newblock {\em Indonesian Journal of Electrical Engineering and Computer Science}, 32(1):517--529, 2023.

\bibitem{reji2023anomaly}
Alan Reji, Bernardi Pranggono, Jims Marchang, and Alex Shenfield.
\newblock Anomaly detection for the internet-of-medical-things.
\newblock In {\em 2023 IEEE International Conference on Communications Workshops (ICC Workshops)}, pages 1944--1949. IEEE, 2023.

\bibitem{robinson2015ranking}
RR~Rejimol Robinson and Ciza Thomas.
\newblock Ranking of machine learning algorithms based on the performance in classifying ddos attacks.
\newblock In {\em 2015 IEEE Recent Advances in Intelligent Computational Systems (RAICS)}, pages 185--190. IEEE, 2015.

\bibitem{roy2022sql}
Prince Roy, Rajneesh Kumar, and Pooja Rani.
\newblock Sql injection attack detection by machine learning classifier.
\newblock In {\em 2022 International Conference on Applied Artificial Intelligence and Computing (ICAAIC)}, pages 394--400. IEEE, 2022.

\bibitem{rozam2023xgboost}
Nadhir~Fachrul Rozam and Mardhani Riasetiawan.
\newblock Xgboost classifier for ddos attack detection in software defined network using sflow protocol.
\newblock {\em International Journal on Advanced Science, Engineering \& Information Technology}, 13(2), 2023.

\bibitem{saed2022detection}
Muhanna Saed and Ahamed Aljuhani.
\newblock Detection of man in the middle attack using machine learning.
\newblock In {\em 2022 2nd International Conference on Computing and Information Technology (ICCIT)}, pages 388--393. IEEE, 2022.

\bibitem{sahu2021malignant}
Laki Sahu, Sanjukta Mohanty, Sunil~K Mohapatra, and Arup~A Acharya.
\newblock Malignant web sites recognition utilizing distinctive machine learning techniques.
\newblock In {\em Computer Networks, Big Data and IoT: Proceedings of ICCBI 2020}, pages 497--506. Springer, 2021.

\bibitem{saini2020detection}
Parvinder~Singh Saini, Sunny Behal, and Sajal Bhatia.
\newblock Detection of ddos attacks using machine learning algorithms.
\newblock In {\em 2020 7th International Conference on Computing for Sustainable Global Development (INDIACom)}, pages 16--21. IEEE, 2020.

\bibitem{schultz2000data}
Matthew~G Schultz, Eleazar Eskin, F~Zadok, and Salvatore~J Stolfo.
\newblock Data mining methods for detection of new malicious executables.
\newblock In {\em Proceedings 2001 IEEE Symposium on Security and Privacy. S\&P 2001}, pages 38--49. IEEE, 2000.

\bibitem{sethi2018novel}
Kamalakanta Sethi, Shankar~Kumar Chaudhary, Bata~Krishan Tripathy, and Padmalochan Bera.
\newblock A novel malware analysis framework for malware detection and classification using machine learning approach.
\newblock In {\em Proceedings of the 19th international conference on distributed computing and networking}, pages 1--4, 2018.

\bibitem{shabtai2012andromaly}
Asaf Shabtai, Uri Kanonov, Yuval Elovici, Chanan Glezer, and Yael Weiss.
\newblock “andromaly”: a behavioral malware detection framework for android devices.
\newblock {\em Journal of Intelligent Information Systems}, 38(1):161--190, 2012.

\bibitem{shhadat2020use}
Ihab Shhadat, Amena Hayajneh, Ziad~A Al-Sharif, et~al.
\newblock The use of machine learning techniques to advance the detection and classification of unknown malware.
\newblock {\em Procedia Computer Science}, 170:917--922, 2020.

\bibitem{sivasangari2021sql}
A~Sivasangari, J~Jyotsna, and K~Pravalika.
\newblock Sql injection attack detection using machine learning algorithm.
\newblock In {\em 2021 5th International Conference on Trends in Electronics and Informatics (ICOEI)}, pages 1166--1169. IEEE, 2021.

\bibitem{suhuan2019android}
Li~Suhuan and Huang Xiaojun.
\newblock Android malware detection based on logistic regression and xgboost.
\newblock In {\em 2019 IEEE 10th International Conference on Software Engineering and Service Science (ICSESS)}, pages 528--532. IEEE, 2019.

\bibitem{sultan2022man}
Ali Bin~Mazhar Sultan, Saqib Mehmood, and Hamza Zahid.
\newblock Man in the middle attack detection for mqtt based iot devices using different machine learning algorithms.
\newblock In {\em 2022 2nd International Conference on Artificial Intelligence (ICAI)}, pages 118--121. IEEE, 2022.

\bibitem{sunday2023phishing}
Seun~Mayowa Sunday.
\newblock Phishing website detection using machine learning: Model development and django integration.
\newblock {\em Journal of Electrical Engineering, Electronics, Control and Computer Science}, 9(3):39--54, 2023.

\bibitem{suresh2011evaluating}
Manjula Suresh and R~Anitha.
\newblock Evaluating machine learning algorithms for detecting ddos attacks.
\newblock In {\em Advances in Network Security and Applications: 4th International Conference, CNSA 2011, Chennai, India, July 15-17, 2011 4}, pages 441--452. Springer, 2011.

\bibitem{thao2017classification}
Tran~Phuong Thao, Akira Yamada, Kosuke Murakami, Jumpei Urakawa, Yukiko Sawaya, and Ayumu Kubota.
\newblock Classification of landing and distribution domains using whois’ text mining.
\newblock In {\em 2017 IEEE Trustcom/BigDataSE/ICESS}, pages 1--8. IEEE, 2017.

\bibitem{toutsop2020monitoring}
Otily Toutsop, Paige Harvey, and Kevin Kornegay.
\newblock Monitoring and detection time optimization of man in the middle attacks using machine learning.
\newblock In {\em 2020 IEEE Applied Imagery Pattern Recognition Workshop (AIPR)}, pages 1--7. IEEE, 2020.

\bibitem{tripathy2020detecting}
Dharitri Tripathy, Rudrarajsinh Gohil, and Talal Halabi.
\newblock Detecting sql injection attacks in cloud saas using machine learning.
\newblock In {\em 2020 IEEE 6th Intl Conference on Big Data Security on Cloud (BigDataSecurity), IEEE Intl Conference on High Performance and Smart Computing,(HPSC) and IEEE Intl Conference on Intelligent Data and Security (IDS)}, pages 145--150. IEEE, 2020.

\bibitem{tuan2020performance}
Tong~Anh Tuan, Hoang~Viet Long, Le~Hoang Son, Raghvendra Kumar, Ishaani Priyadarshini, and Nguyen Thi~Kim Son.
\newblock Performance evaluation of botnet ddos attack detection using machine learning.
\newblock {\em Evolutionary Intelligence}, 13:283--294, 2020.

\bibitem{udayakumar2018malware}
N~Udayakumar, Vatsal~J Saglani, Aayush~V Cupta, and T~Subbulakshmi.
\newblock Malware classification using machine learning algorithms.
\newblock In {\em 2018 2nd International Conference on Trends in Electronics and Informatics (ICOEI)}, pages 1--9. IEEE, 2018.

\bibitem{ussatova2024development}
Olga Ussatova, Aidana Zhumabekova, Vladislav Karyukin, Eric~T Matson, and Nikita Ussatov.
\newblock The development of a model for the threat detection system with the use of machine learning and neural network methods.
\newblock {\em International Journal of Innovative Research and Scientific Studies}, 7(3):863--877, 2024.

\bibitem{vanitha2019malicious}
N~Vanitha and V~Vinodhini.
\newblock Malicious-url detection using logistic regression technique.
\newblock {\em International Journal of Engineering and Management Research (IJEMR)}, 9(6):108--113, 2019.

\bibitem{vijayalakshmi2023detection}
A~Vijayalakshmi, J~Broody, J~Sri Sathishkumar, et~al.
\newblock Detection of man in the middle attack in 5g iot using machine learning.
\newblock In {\em 2023 International Conference on Recent Advances in Science and Engineering Technology (ICRASET)}, pages 1--5. IEEE, 2023.

\bibitem{wang2023combination}
Zhanfeng Wang, Lisha Yao, Xiaoyu Shao, and Honghai Wang.
\newblock A combination of textcnn model and bayesian classifier for microblog sentiment analysis.
\newblock {\em Journal of Combinatorial Optimization}, 45(4):109, 2023.

\bibitem{xie2019sql}
Xin Xie, Chunhui Ren, Yusheng Fu, Jie Xu, and Jinhong Guo.
\newblock Sql injection detection for web applications based on elastic-pooling cnn.
\newblock {\em IEEE Access}, 7:151475--151481, 2019.

\bibitem{yamashita2018convolutional}
Rikiya Yamashita, Mizuho Nishio, Richard Kinh~Gian Do, and Kaori Togashi.
\newblock Convolutional neural networks: an overview and application in radiology.
\newblock {\em Insights into imaging}, 9:611--629, 2018.

\bibitem{yaswanth2023prediction}
Palla Yaswanth and V~Nagaraju.
\newblock Prediction of phishing sites in network using naive bayes compared over random forest with improved accuracy.
\newblock In {\em 2023 Eighth International Conference on Science Technology Engineering and Mathematics (ICONSTEM)}, pages 1--5. IEEE, 2023.

\bibitem{yusof2016evaluation}
Ahmad~Riza’ain Yusof, Nur~Izura Udzir, and Ali Selamat.
\newblock An evaluation on knn-svm algorithm for detection and prediction of ddos attack.
\newblock In {\em Trends in Applied Knowledge-Based Systems and Data Science: 29th International Conference on Industrial Engineering and Other Applications of Applied Intelligent Systems, IEA/AIE 2016, Morioka, Japan, August 2-4, 2016, Proceedings 29}, pages 95--102. Springer, 2016.

\bibitem{zekri2017ddos}
Marwane Zekri, Said El~Kafhali, Noureddine Aboutabit, and Youssef Saadi.
\newblock Ddos attack detection using machine learning techniques in cloud computing environments.
\newblock In {\em 2017 3rd international conference of cloud computing technologies and applications (CloudTech)}, pages 1--7. IEEE, 2017.

\bibitem{zhu2020dtof}
Erzhou Zhu, Yinyin Ju, Zhile Chen, Feng Liu, and Xianyong Fang.
\newblock Dtof-ann: an artificial neural network phishing detection model based on decision tree and optimal features.
\newblock {\em Applied Soft Computing}, 95:106505, 2020.

\end{thebibliography}

\end{document}